\documentclass[twoside]{class-definitions}
\usepackage[latin1]{inputenc}
\usepackage{wrapfig,rotating}
\usepackage{amssymb,amsmath,array}

\newcommand{\Wqq} {$W \rightarrow q\bar{q^{\prime}}$ }
\newcommand{\Zqq} {$Z \rightarrow q\bar{q}$ }
\newcommand{\Wev} {$W \rightarrow e \nu$ }
\newcommand{\Wuv} {$W \rightarrow \mu \nu$ }

\newcommand{\Zuu} {$Z \rightarrow \mu^+ \mu^-$ }
\newcommand{\dream} {{\sc dream }}
\newcommand{\C}    {\^{C}erenkov }

\begin{document}
\title{Another Detector \\  for the International Linear Collider} 
\author{Nural Akchurin$^1$, Sehwook Lee$^1$, Richard Wigmans$^1$, Hanna Arnold$^2$, \\ Aaron Bazal$^2$,   Robert Basili$^2$,  John Hauptman$^2$,   Tim Overton$^2$, \\ Andrew Priest$^2$,  Bingzhe Zhao$^2$,  Alexander Mikhailichenko$^3$,  \\ Michele Cascella$^4$, Franco Grancagnolo$^4$,  Giovanni Tassielli$^4$, \\  Franco Bedeschi$^5$,  Fabrizio Scuri$^5$,  Sung Keun Park$^6$,  \\
 Fedor Ignatov$^7$,  Gabriella Gaudio$^9$, Michele Livan$^8$
\thanks{This work has been supported by the US Department of Energy and by INFN, Italy.}
\vspace{.3cm}\\
1 - {\it Texas Tech University, Department of Physics}, Lubbock, TX 79409-1051 USA \\
2 - {\it Iowa State University, Department of Physics and Astronomy},
Ames, IA  50011  USA \\
3- {\it Laboratory for Elementary Particle Physics, Cornell University}, Ithaca, NY 14853-5001 USA \\
4 - {\it INFN and Dipartimento di Fisica},  via Lecce-Arnesano, 73100,   Lecce, Italy \\
5 - {\it INFN, Sezione di Pisa}, via Livornese 582/a, I-56010 S. Piero a Grado, Pisa, Italy \\
6 - {\it Korea University, Department of Physics}, Anam-dong, Seoul 136-701, Korea \\
7 - {\it Budker Institute of Nuclear Physics}, Lavrentyeva-11, RU-630 090 Novosibirsk, Russia  \\
8 - {\it Dipartimento di Fisica, Universita di Pavia}, Italy \\
9- {\it INFN Sezione di Pavia}, Italy
} 

\maketitle



\bigskip \bigskip \centerline{\bf Abstract} \bigskip

\begin{abstract}
We describe another detector\footnote{We can generate names and acronyms, but we do not have a name:  {\sc audax}, audacious, bold ({\it Latin}); CAD, Compact Advanced Detector; or, CLD, Compact Lightweight Detector, {\it etc.}.}  designed for the International 
Linear Collider based on several tested instrumentation innovations in order
to achieve the necessary experimental goal of a detecter that is 2-to-10 
times better than the already excellent SLC and LEP detectors, in particular, (1) 
dual-readout calorimeter system based on the RD52/DREAM measurements at CERN, 
(2) a cluster-counting drift chamber based on the successful {\sc kloe} chamber at Frascati, 
and (3) a second solenoid to return the magnetic flux without iron.   A high-performance pixel
vertex chamber is presently undefined.  We discuss
particle identification, momentum and energy resolutions, and the machine-detector
interface that together offer the possibility of a very high-performance detector for
$e^+e^-$physics up to $\sqrt{s} = 1$ TeV.  
\end{abstract}


\newpage
\section{Introduction}

The physics reach of  a new  high energy $e^+e^-$ linear  collider requires \cite{DCR} the 
measurement and identification of all known partons of the standard model 
($e, \mu, \tau, \nu, uds, c, b, t)$ and the bosons $(W, Z, \gamma,$ and $H^0$)  including the hadronic decays of the
gauge bosons, \Wqq and \Zqq and, by subtraction, the missing neutrinos in \Wev, \Wuv,
$\tau \rightarrow \ell \nu_{\ell} \nu_{\tau}$ and $\tau \rightarrow h \nu_{\tau}$  decays
so that kinematically over-constrained final states can be achieved.  

An important benchmark process is
\begin{displaymath}
 e^+e^- \rightarrow H^0 Z^0 \rightarrow (anything) + \mu^+\mu^-
\end{displaymath}
in which the two $\mu$s are measured in the tracking system and the Higgs is
seen in the missing mass distribution against the $\mu^+\mu^-$ system.  This decay mode 
stresses the tracking system.   A
momentum resolution of $\sigma_p/p^2 \approx 4 \times 10^{-5}$ (GeV/c)$^{-1}$
is required for a desired  Higgs mass resolution of 150 MeV/c$^2$ in a 500 fb$^{-1}$
data sample.  There are
three main technologies under study to achieve this performance:
a 5-layer silicon strip tracker ({\sc sid}), a TPC with sophisticated high-precision end 
planes ({\sc ild}), and a cluster-counting drift chamber (this detector) \cite{Trk-Review}.

This same final state can be studied for \Zqq  decays 
\begin{displaymath}
 e^+e^- \rightarrow H^0 Z^0 \rightarrow (anything) + q+\bar{q}  \rightarrow (anything) + {\rm jet + jet}
\end{displaymath}
which are 20 times more plentiful than \Zuu decays but less distinct experimentally.  
This mode stresses the hadronic  calorimeters.   In addition, those 
processes that produce $W$ and $Z$ bosons either by production 
$(e^+e^- \rightarrow W^+W^-, Z^0Z^0, HHZ)$ or by decay $(H \rightarrow W^+W^-)$,
will rely critically on the direct mass resolution on \Zqq and \Wqq decays, and
these processes demand that the calorimeter energy
resolution be as good as $\sigma_E/E \approx 30\%/\sqrt{E}$ with a constant term less
than 1\%.  There are two main technologies under study:  
a highly segmented calorimeter volume with approximately (1 cm)$^3$ channels
for the implementation of Particle Flow Analysis (PFA) algorithms ({\sc sid} and {\sc ild}), and dual-readout optical calorimeters that measure both scintillation and Cerenkov light (this detector) \cite{Cal-Review}.

\begin{wrapfigure}{r}{0.7\columnwidth} 
\centerline{\includegraphics[width=0.69\columnwidth]{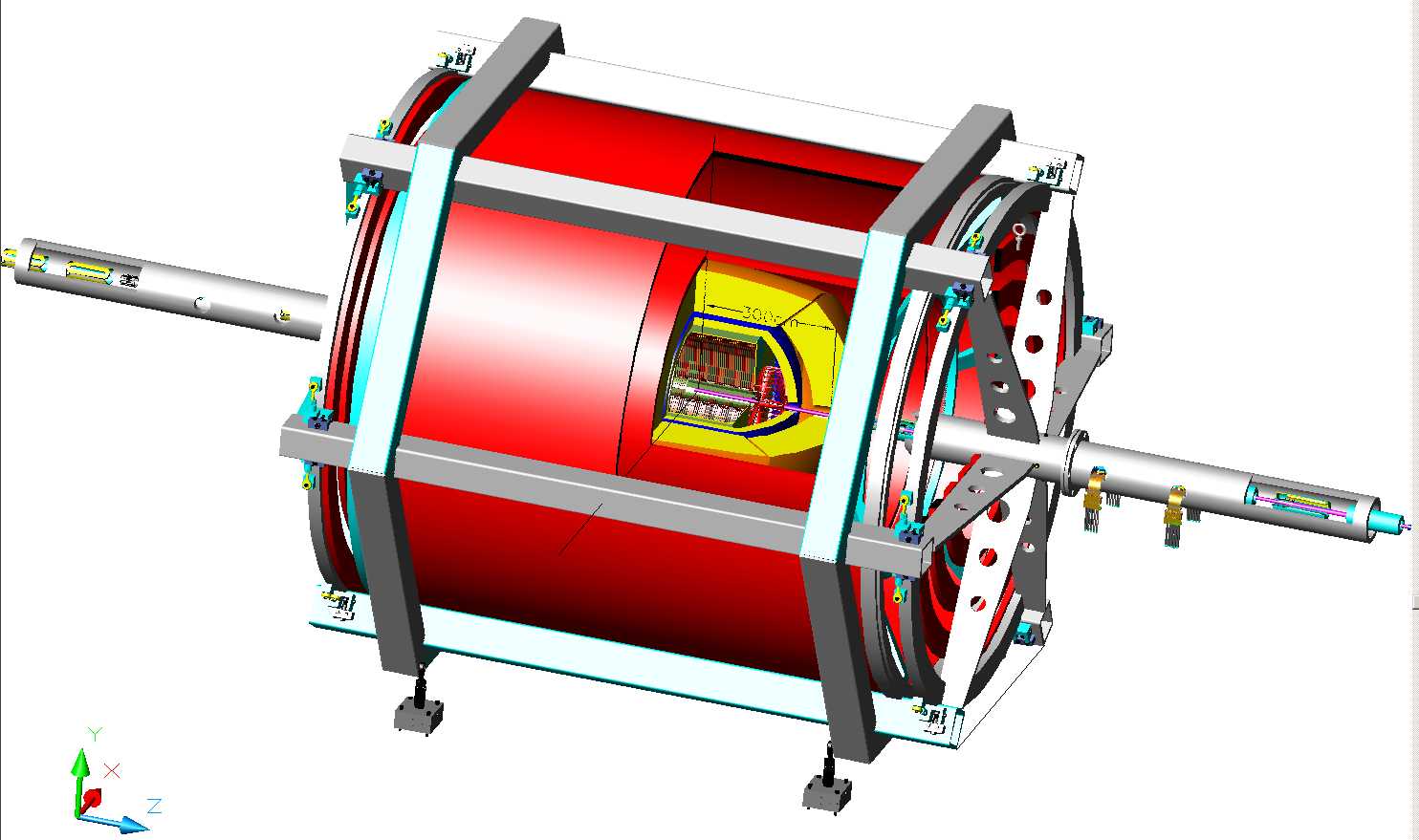}}
\caption{This alternative detector.  Dual-solenoids in red; dual-readout calorimeter in yellow; cluster-timing tracking chamber  in center; and machine-detector final focus elements in grey.}\label{fig:4th}
\end{wrapfigure}

The identification of $b$ and $c$ quark and $\tau$ lepton decays is critical
to good physics since these massive particles are a gateway to the decays of
more massive, and possibly new, particles \cite{Vtx-Review}.   The measurement of 
their respective decay lengths places stringent conditions on the spatial
precision of a vertex chamber and how close it can be to the beam interaction
spot.  Typically, a spatial resolution of $\sigma \approx 5 ~\mu$m is desired,
with large solid angular coverage.   The inner radius is limited by the debris
from beamstrahlung that is only suppressed by the axial tracking magnetic
field; the uncharged debris cannot be suppressed, but becomes nearly invisible in
our He-based tracking system which presents a negligibly thin absorber for the
$\gamma$'s from the violent beam crossing.

The region beyond the calorimeter is reserved for, typically, a superconducting solenoid
to establish the tracking field and a hadron absorber
or muon filter to assist muon tagging.   There are two main types of muon systems
under study:  an iron absorber interspersed with tracking chambers to measure
the trajectories of penetrating tracks  ({\sc sid} and {\sc ild}), and an iron-free design in which the magnetic flux is returned by a second outer solenoid (this detector).

We are skeptical of several aspects of the  {\sc sid} and {\sc ild} detectors, {\it viz.}, power pulsing of a complex silicon tracking system, the analysis complexity of a particle flow calorimeter, positive ion loading in a TPC, and a general lack of particle identification measurements.

\section{An Alternative Detector}

\begin{wrapfigure}{r}{0.55\columnwidth} 
\centerline{\includegraphics[width=0.55\columnwidth]{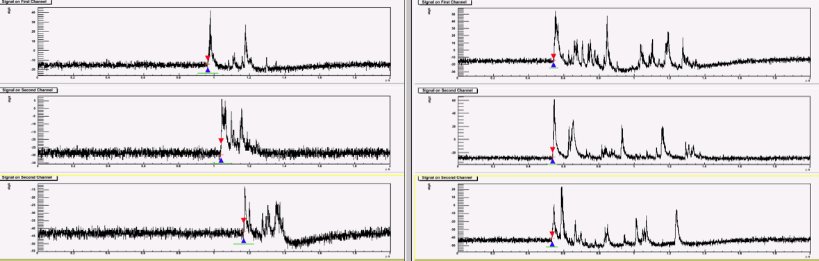}}
\caption{Ionization cluster data.}\label{fig:clusters}
\end{wrapfigure}

We have introduced new ideas and instruments in order to achieve the resolution
requirements needed for ILC physics studies \cite{DOD,LoI}.  With the exception of the vertex
chamber, we designed a powerful detector that dramatically differs from the detectors 
at SLC and LEP and from the two
ILC concept detectors, {\sc ild} and {\sc sid}, in all three major
detector systems:  the tracking, the calorimeter and the muon system.  This detector
is displayed in Figure \ref{fig:4th}, showing the beam transport through the final focus,
the dual solenoids (red),  fiber dual-readout calorimeters (yellow), the
tracking chamber, and the vertex chamber.   The muon tracking is in the annulus
between the solenoids.  A forward toroid for high precision forward tracking in under
current study in this figure.  This detector is about 1/10 the mass of a 
conventional detector with an iron yoke flux return.

\subsection{Tracking by cluster counting in a low-mass He-based drift chamber}

The gaseous central tracker is a cluster-counting drift chamber
modeled on the successful {\sc kloe} main tracking chamber  \cite{Trk-Review,LoI,pisa}.  This drift chamber 
(CluCou) maintains very low multiple scattering due to a He-based gas and aluminum 
wires in the tracking volume and  utilizes a particular assembly scheme, which makes use of 
an ultra-light wire support, developed for the Mu2e I-Tracker\footnote{The same scheme is being implemented for the MEG upgrade drift chamber and for the tracking detector at the proposed tau/charm Factory in Italy.} Forward tracks 
(beyond $\cos \theta \approx 0.7$) which  penetrate the wire support frame and 
electronics pass through less than 2\% $X_{0}$ of material.   The ultra-low mass of 
the tracker in the central region directly improves momentum resolution in the multiple scattering dominated region 
below 10 GeV/c.  The He gas has a low drift velocity which allows a new cluster/timing 
 technique that clocks in individual ionization clusters on every 
wire, providing an estimated 60 micron spatial resolution per point,  particle identification, trough cluster counting, at least a factor 2 better than the resolution attainable with $dE/dx$,  $z$-coordinate information on each track segment through an effective dip 
angle measurement,  and a layout made exclusively of super-layers with alternated opposite sign stereo angles.  The maximum drift time in each cell is less than the 300ns beam crossing 
interval, so this chamber sees  only one crossing per readout. Furthermore, the high sampling rate ($>$ 2GSa/s) and the high multiplicity of hits per track ($>$200) allow for the determination of the event time (trigger time) to better than a fraction of a ns. Figure \ref{fig:clusters}   (attached below) shows two cosmic ray track segments through a three drift tubes (2 cm radius) chamber prototype. The one on the right passes close to the sense wire plane and exhibits a large number of clusters, the one on the left is a large angle track segment passing far from the sense wires, with a smaller number of clusters due to the shorter path length. The horizontal full scale is 2 microseconds.

\begin{wrapfigure}{r}{0.55\columnwidth} 
\centerline{\includegraphics[width=0.50\columnwidth]{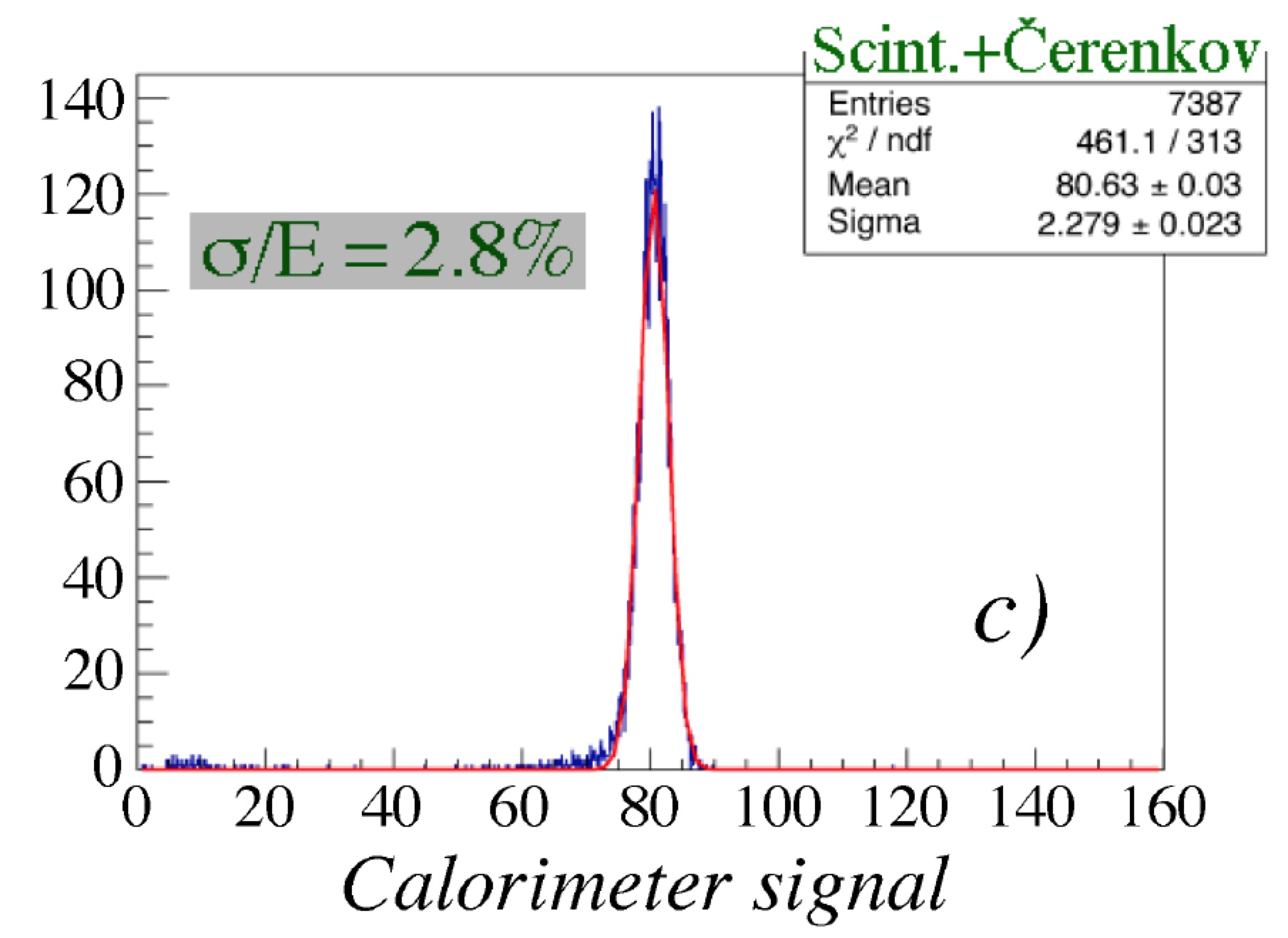}}
\caption{RD52 energy resolution for 40 GeV electrons.  This is the {\it sum} of the $S$ and $C$ distributions.\cite{RD52-EM}}\label{Fig:EM-resol}
\end{wrapfigure}

The critical issues of occupancy and two-track resolution have been simulated for ILC 
events and expected machine and event backgrounds, and direct GHz cluster counting 
experiments are being performed.  This chamber has timing and pattern recognition
capabilities midway between the faster, higher precision silicon tracker of {\sc sid}  and the slower, 
full 3-dimensional imaging TPC of {\sc ild}, and is superior to both with respect to its low multiple scattering.

The low-mass of the tracking medium, the multiplicity of point measurements (about 200), and the point
spatial precision allow this chamber to reach $\sigma_p/p^2 \approx 5 \times 10^{-5}$ (GeV/c)$^{-1}$
at high momenta, and to maintain good momentum resolution down to low momenta.

\subsection{Calorimetry by dual-readout of scintillation and \C light}

\begin{figure}[t!]
\includegraphics[width=5.5in]{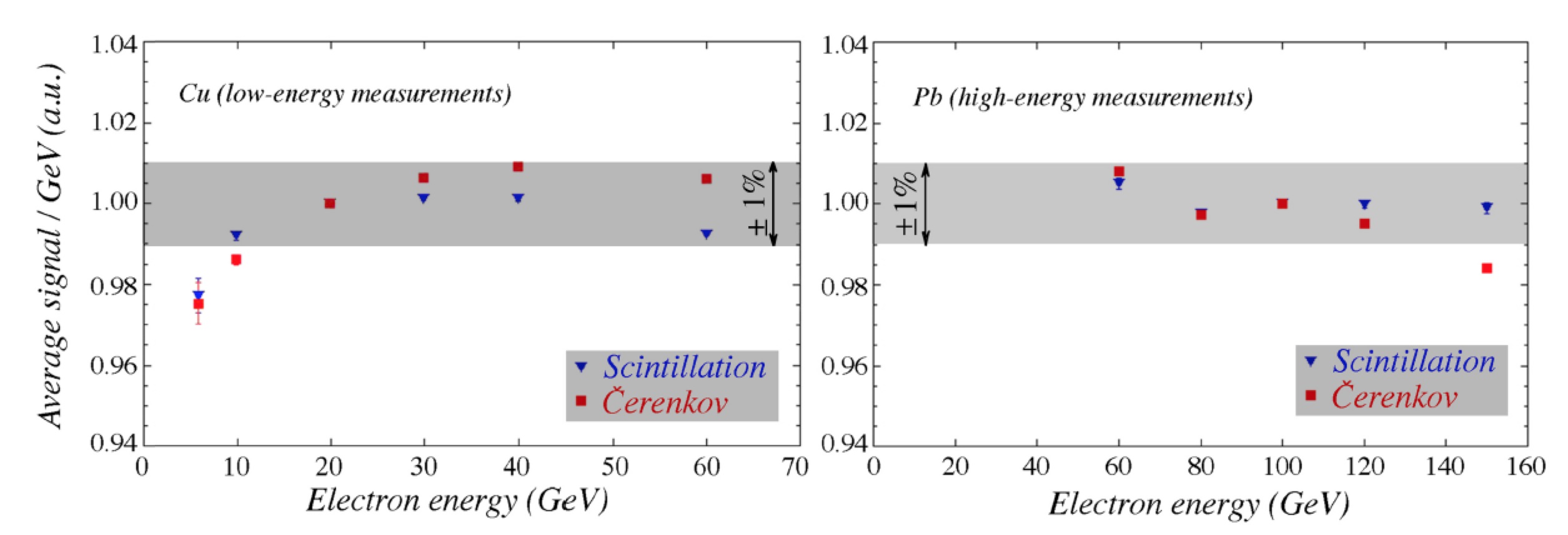}
\caption{Measured response of the dual readout calorimeters of RD52 for
  hadrons from 5 to 150 GeV.  Each module was calibrated
  only with 60 GeV electrons.  The 2\% drop at low beam energy is due to material in the upstream detectors such as the  trigger counters, wire chambers, and backscattering at the front face of the calorimeter.}\label{Fig:linearity}
\end{figure}

The calorimeter is a spatially fine-grained dual-readout fiber
sampling calorimeter augmented with the ability to measure the 
neutron content of a shower\cite{LoI}.  The dual fibers are sensitive to scintillation and 
Cerenkov radiation, respectively, for separation of the hadronic and electromagnetic components of 
hadronic showers \cite{dream-ha}, and since fluctuations in the {\sc em} fraction, i.e.,
fluctuations in $\pi^0$ production relative to $\pi^{\pm}$, are largely responsible for poor hadronic energy
resolution, the \dream module achieved a substantial improvement in hadronic
calorimetry.  The energy
resolution of the tested \dream calorimeter has been surpassed in the new RD52 modules
 with finer spatial sampling, neutron detection for the measurement of fluctuations
in binding energy losses\cite{dream-n}, and a larger volume test module to reduce
leakage fluctuations.  The calorimeter modules
will have fibers up to their edges, and will be constructed for sub-millimeter
close packing, with signal extraction at the outer radius so that the
calorimeter system will approach full coverage without cracks.  The electromagnetic
energy resolution, spatial resolution, and particle identification are so good in the fiber system than a 
separate crystal dual-readout in not desired.

\begin{figure}[b!]
 \centerline{\includegraphics[width=14.cm]{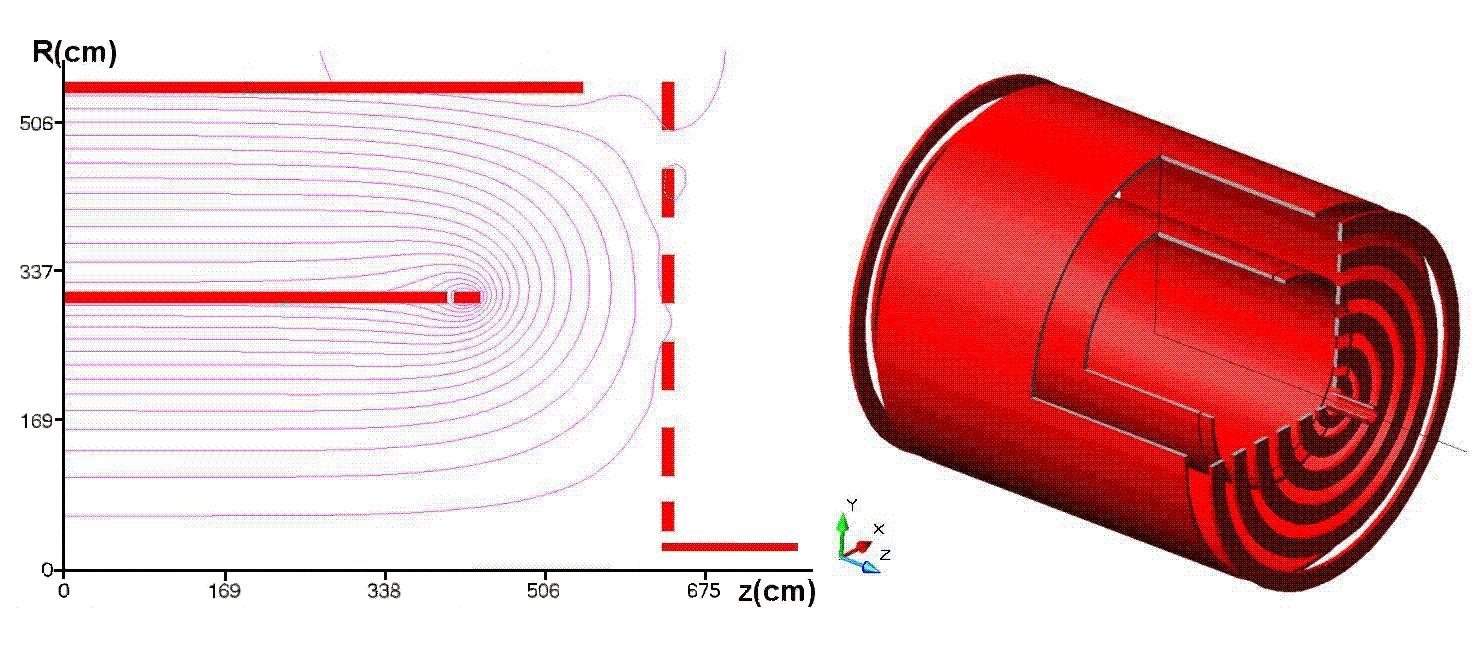}}
 \caption{Drawings   showing the two solenoids and the ``wall of coils''
   that redirects the field out radially, and the resulting field lines in an
     $r-z$ view.  This field is uniform to 1\% at $3.5$ T in the tracking
     region, and also uniform and smooth at $-1.3$ T in the muon
     tracking annulus between the solenoids. }
 \label{fig:B+coils}
\end{figure}

The fiber calorimeter shows promise of excellent energy resolution on
hadrons and jets, as detailed in recent progress reports \cite{SPSC}.  We expect to achieve
a hadronic energy resolution near 1-2\% on a full-sized calorimeter,
\begin{displaymath}
 \frac{\sigma}{E} \approx 1-2 \%  ~~~~~ {\rm (hadronic~ energy~ resolution~ on~ jets)}
\end{displaymath}
at high energies, including a small constant term (which has many sources).

Most importantly, the hadronic response of this 
dual-readout calorimeter is demonstrated to be linear in hadronic
energy from 20 to 300 GeV having been {\it calibrated with electrons at only one energy}
 Fig. \ref{Fig:linearity}. This is a critical advantage at the ILC where
calibration with 45 GeV electrons from $Z$ decay will suffice to maintain the energy
scale up to 10 times this energy for physics.

Finally, we expect to construct a small tungsten dual-readout module identical in geometry to our copper and lead modules.  Such a development would retain the good features of dual-readout but reduce the calorimeter depth to 1.5 meters, resulting in enormous economies in a large $4\pi$ collider detector.


\begin{wrapfigure}{r}{0.6\columnwidth}
  \includegraphics[width=7.5cm,height=9.cm]{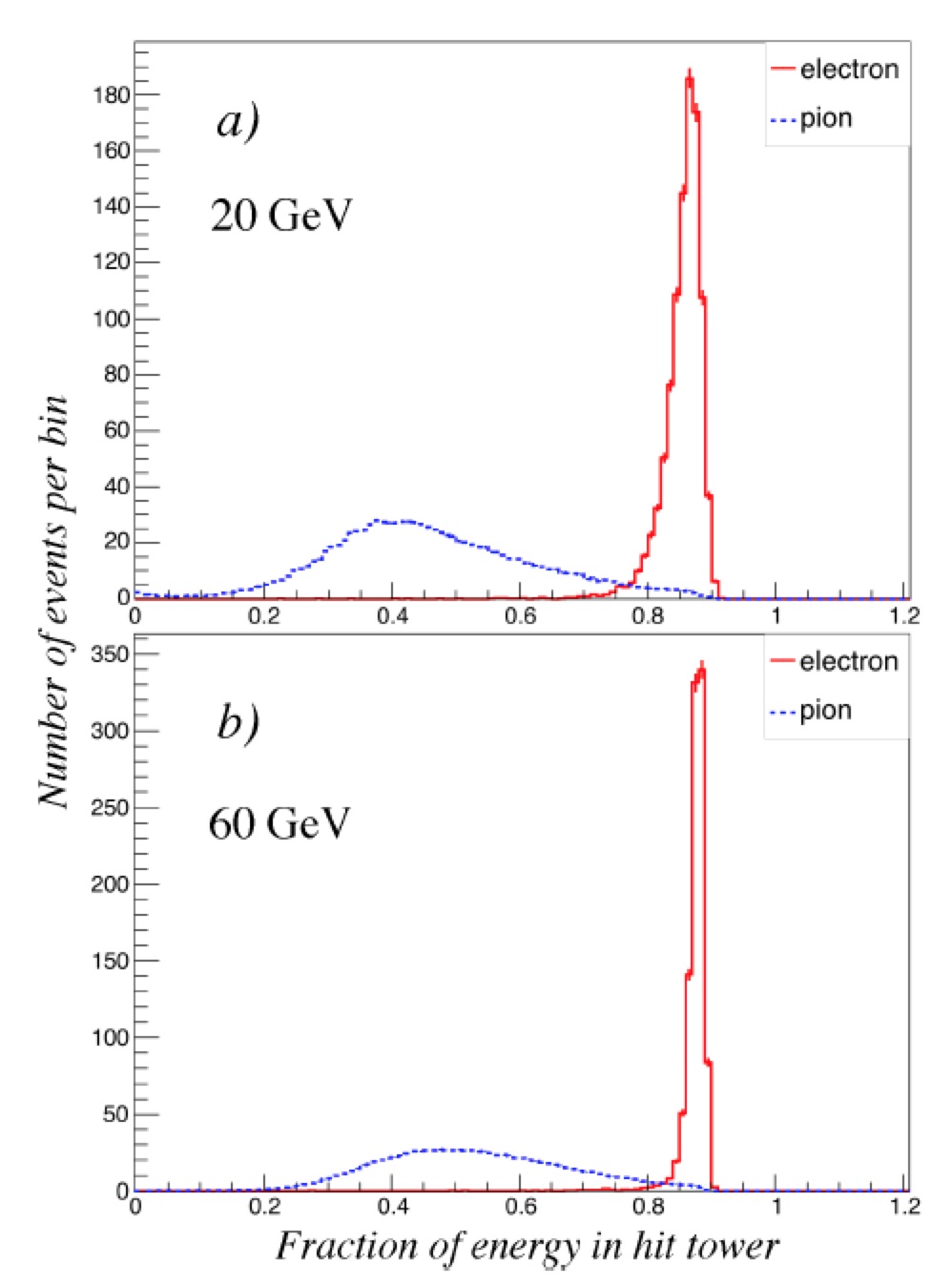}
  \caption{(a) The fraction of energy deposited in the hit tower for 20 GeV $\pi^-$'s and $e^-$'s;  (b) same for 60 GeV $\pi^-$'s and $e^-$'s.    }
  \label{fig:e-pi}
\end{wrapfigure}

\subsection{The  dual solenoid magnetic field configuration}

The muon system utilizes a dual-solenoid magnetic field configuration 
in which the flux from the inner solenoid is returned
through the annulus between this inner solenoid and an 
outer solenoid oppositely driven with a smaller turn density \cite{Mik}.  
The magnetic field in the volume between 
the two solenoids will back-bend muons (and punch-through pions) which have penetrated the calorimeter 
and allow, with the addition of tracking chambers, a second momentum measurement.
This will achieve high precision without the limitation of 
multiple scattering in $Fe$ that fundamentally limits
momentum resolution in conventional muon systems to 10\%.    
High spatial precision drift tubes with cluster counting
electronics are used to measure tracks in this volume  \cite{franco-mu-val}.  
The dual-solenoid field is terminated by a novel ``wall of coils''
that provides muon bending down to small angles 
($\cos \theta \approx 0.975$) and also
allows good control of the magnetic environment 
on and near the beam line. The design is illustrated in Fig~\ref{fig:B+coils}.

The path integral of the field in the annulus for  a muon from 
the origin is about 3 T$\cdot$m
over $0 < \cos \theta < 0.85$ and remains larger than 0.5 T$\cdot$m
out to $\cos \theta = 0.975$, allowing both good momentum resolution
and low-momentum acceptance over almost all of $4\pi$ \cite{franco-mu-val}.

\begin{wrapfigure}{r}{0.6\columnwidth}
\centerline{\includegraphics[width=0.55\columnwidth]{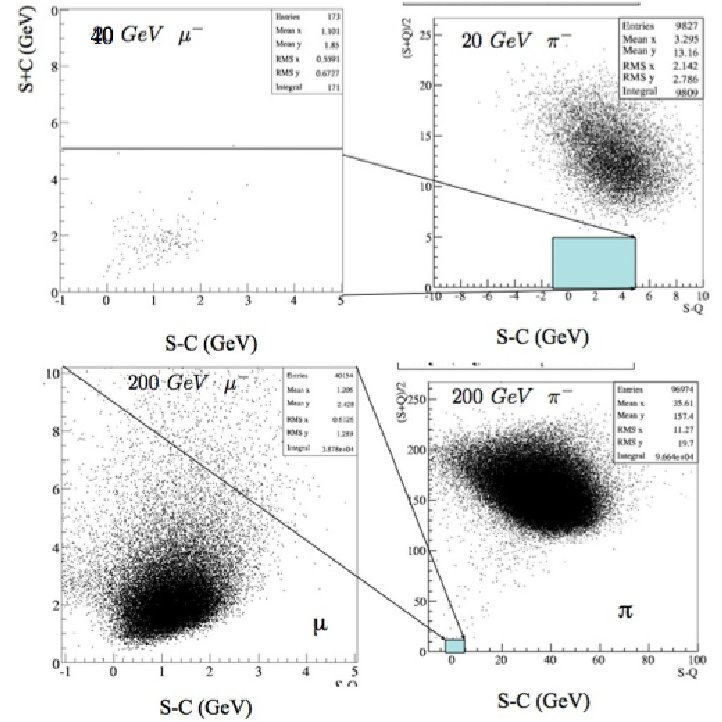}}
\caption{Dual-readout separation of $\mu-\pi^{\pm}$.}\label{fig:mu-pi}
\end{wrapfigure}

For isolated tracks, the dual readout calorimeter independently provides a unique 
identification of muons relative to pions with a background track
rejection of $10^4$, or better, for high energies through its separate measurements
of ionization and radiative energy losses, Fig. \ref{fig:mu-pi}.  This will be important 
for the $\tau^{\pm}$ decays from $H^0$ and $W,Z$ leptonic decays.

\section{Particle Identification}

The capability to identify standard model  partons ($e, \mu, \tau, uds,$
 $c, b, t, W, Z, \gamma$) is equivalent to increased luminosity with larger,
and less ambiguous, data ensembles available for physics analysis.

\begin{wrapfigure}{l}{0.6\columnwidth}
  \includegraphics[width=7.cm,height=5.cm]{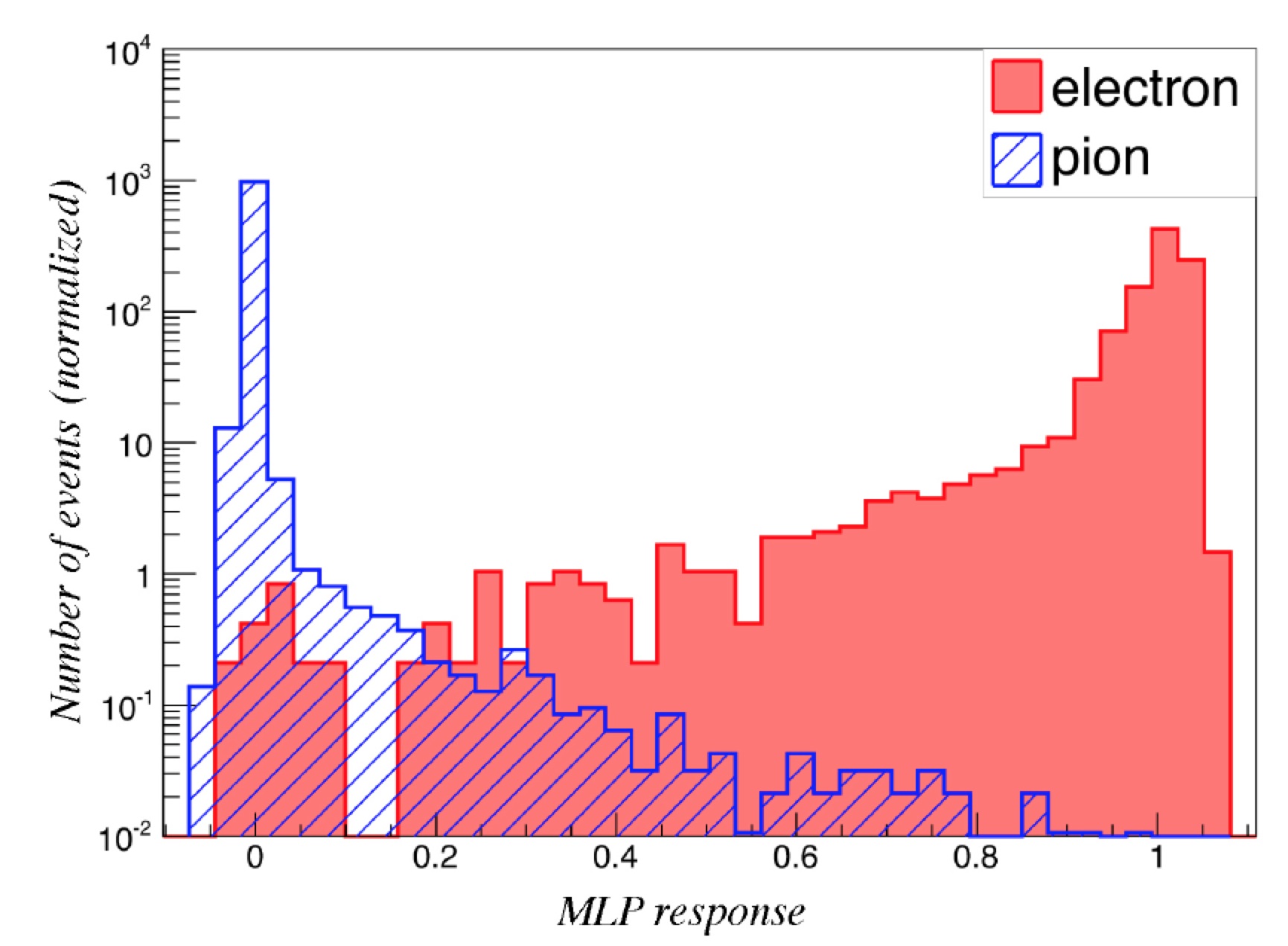}
  \caption{A multi-variate discriminant, MLP, of $e^{\pm}$ from $\pi^{\pm}$ using the fraction 
  of Fig. \ref{fig:e-pi}, an S/C measurement, and a time-depth measurement.  $\pi^{\pm}$ rejections of 99.9\% for $e^{\pm}$ efficiencies of 99.9\% are achievable.}
  \label{fig:MLP}
\end{wrapfigure}

\paragraph{$\pi^0 \rightarrow \gamma \gamma$ separation and reconstruction}

The dual-readout crystals can be made small, about 1cm$\times$1cm or 2cm$\times$2cm,
and with reconstruction using shower shapes in the crystals we estimate that 
$\pi^0 \rightarrow \gamma \gamma$ can be reconstructed up to about $E_{\pi^0} \sim 20$ GeV,
which is high enough to reconstruct the important decay 
  $\tau \rightarrow \rho \nu \rightarrow \pi^{\pm} \pi^0 \nu \rightarrow \pi^{\pm} \gamma \gamma \nu$
to be used as a spin analyzer in the decays of the 125 GeV/c$^2$ $H^0$, etc.

\paragraph{$e, \pi, K, p$ separation by $dE/dx$ at lower momenta}

The cluster counting central tracking chamber has the added benefit of an
excellent energy loss measurement without a Landau tail since clusters are
counted as Poisson.    We anticipate 3\% or
better resolution in $dE/dx$ as an analysis tool in, for example,  $b$ physics
where a large fraction  of charged tracks are below a few GeV/c.

 \paragraph{$e$ separation from $\pi^{\pm}$ and jets $(j)$}


Dual-readout allows many measurements \cite{spacal} to discriminate $e^{\pm}$ from $\pi^{\pm}$.  A lateral
shape discriminant \cite{RD52-pID}  in Fig. \ref{fig:e-pi} is very effective for narrow fiber channels; when combined with depth-time information and a scintillation-to-\C measurement, discrimination reaches levels of 1000-to-1, Fig. \ref{fig:MLP}. 

Direct use of the scintillation {\it vs.} the \C responses, both for the overall shower and in individual channels, provides further discrimination:  just as the overall shower   $C ~vs.~ S$ response 
fluctuates in hadronic showers, so do the individual channels of $\pi^{\pm}$ showers.  The statistic
 \begin{displaymath}
 \sigma_{Q-S} = \frac{1}{N}  \sum_{i=1}^{N} (Q_i - S_i)^2
 \end{displaymath}
 is a measure of these channel-to-channel fluctuations.  
 For 100 GeV $e$ this $\sigma^e \approx 0.2$ GeV$^2$, and 
 for $\pi^{\pm}$ $\sigma^{\pi} \approx 10$ GeV$^2$, yielding a rejection of
 $\pi^{\pm}$ against $e$ of about 50.

   The time history of the 
   scintillating signal contains independent information, in particular, the neutrons
   generated in the $\pi^{\pm}$ cascade travel slower ($v \sim 0.05 c$) and fill a
   larger volume, and therefore the elapsed time of the scintillating signal is longer
   for a $\pi^{\pm}$ and an $e$, shown by the {\sc spacial} calorimeter \cite{spacal} to achieve 
   a discrimination of about 100-to-1.

     The exploitation of these measurements in 
   a collider experiment will depend on many factors, such as channel size, the 
character of event ensembles, and the fidelity of the measurements themselves.  
The goal of this alternative detector is to package these capabilities into   a comprehensive
detector.
  

\paragraph{$\mu$ separation from $\pi^{\pm}$}

We have achieved excellent $\mu-\pi^{\pm}$ separation in the dual
readout calorimeter and additional separation using energy balance
from the tracker through the calorimeter into the muon spectrometer.
A non-radiating $\mu$ has a zero \C signal in the fiber calorimeter since
the \C angle is larger than the capture cone angle of the fiber.  The 
scintillating fibers measure $dE/dx$ of the through-going $\mu$ \cite{dream-mu}.
Any radiation by the $\mu$ within the body of the calorimeter is sampled
equally by the scintillating ($S$) and \C ($C$) fibers \cite{dream-e}, and therefore
$S-C \approx dE/dx$ independent of the amount of radiation.  The distributions of $(S-C)$
 {\it vs.} $(S+C)/2$ for 20 GeV and 200 GeV $\pi^-$ and $\mu^-$ are shown in Fig.
 \ref{Fig:mu-pi} in which it is evident that for an isolated track the $\pi^{\pm}$ 
 rejection against $\mu$ is about $10^4$ at 20 GeV and $10^5$ at 200 GeV.
 A further factor of 50 is obtained from the iron-free dual solenoid in which 
 the precisely measured $\mu$ momentum can be matched with the momentum
 in the central tracker and the radiated energy in the calorimeter. 
  We expect that other effects, such as tracking inefficiencies, will limit this level of
 rejection before these beam-measured rejections are achieved in practice.

\paragraph{Time-of-flight in CluCou cluster-timing  and the \C fibers}

The time history readout of the scintillating and \C fibers will serve several purposes,
{\it viz.}, $e-\pi$ separation (Fig. \ref{fig:e-pi}(b)), neutron measurement for the
suppression of fluctuations in  binding energy losses, and as a real-time
nanosecond monitor of all activity in the volume between the 337-ns bunch
crossings, including 'flyers' and beam burps of any kind.

In addition, the sub-$ns$ resolution on the time of arrival of a
shower can be used in the time-of-flight of heavy objects such as
supersymmetric or technicolor particles that move slowly ($v \sim 0.1c$) 
into the tracking volume, where the cluster-timing is also sub-$ns$, 
 and decay to light particles ($e, \mu, \tau, j$, etc.).  Such objects 
can be easily reconstructed in this detector. 

\section{Machine-Detector Interface}

The detector's magnetic field is confined essentially to a cylinder with
negligible fringe fields, without the use of iron flux return, Fig. \ref{fig:B+coils} . 
This scheme offers flexibility in controlling the fields 
along the beam axis and, in particular, making the fields on the delicate final focus zero, 
or very close to zero.  The twist compensation solenoid
just outside the wall of coils is shown in Fig. \ref{fig:4th}, along
with the beam line elements close to the IP.  This iron-free 
configuration \cite{Mik}
allows us to mount all beam line elements on a single support and
drastically reduce the effect of vibrations at the final focus (FF),
essentially because the beams will coherently move up and down together.
In addition, the FF elements can be brought closer to the vertex chamber
for better control of the beam crossing.

The open magnetic geometry of this detector also allows for future 
physics flexibility for asymmetric energy collisions, the installation
of specialized detectors outside the inner solenoid, and
magnetic flexibility for non-zero dispersion FF optics at the IP, 
adiabatic focussing at the IP, and monochromatization of the
collisions to achieve a minimum energy spread \cite{Mik}.
Finally, this flexibility and openness allows additions
in later years  to the detector or to the beam line, and therefore no
physics  is precluded by this detector concept.  A prime example is
$\gamma$-$\gamma$ physics \cite{telnov}, but there are many more 
exotic possibilities in long-lived dark matter particles, massive slow-moving
SUSY particles, {\it etc.},  that would remain unseen in a conventional detector.

\section{Summary}

This alternative detector contains many new ideas in high energy physics
instrumentation that are aimed at a comprehensive detector up to 1 TeV $e^+e^-$
physics at the International Linear Collider.  All of the data shown in this paper
and all of the performance specifications have already been tested in either
beam tests, prototypes, or existing detectors.  The difficult problems of 
incorporating these small successful instruments into a large detector 
while maintaining the scientific strengths of each present good work  in
the near future.







\begin{thebibliography}{99}
\bibitem{DCR} International Linear Collider Reference Design Report (2007) Vol. 4,
Detectors, August 2007, Eds. Chris Damerell, John Jaros, Akiya Miyamoto, and Ties Behnke; and, 
{\tt http://www.\-linearcollider\-.org/\-ILC/\-Publications/\-Technical-\-Design-\-Report}, (2013).

\bibitem{Trk-Review} ILC Tracking R\&D: Report of Review Committee, 
Beijing, China, 5-8 Feb 2007, (BILCW07), 
   Chris Damerelll (Chair), D. Karlen, W. Lohmann, H. Park, H. Weerts, 
   P. Braun-Munzinger, I. Giomataris, H. Hamagaki, F. Sauli, H. Spieler, 
   M. Tyndel, Y. Unno, C. Yuanbo, O. Qun. J. Brau, J. Haba, B. Zhou
   15 April 2007.
  
     
\bibitem{Cal-Review}     ILC Calorimetry R\&D Review, DESY ILC Workshop,
31 May - 4 June 2007, W. Lohmann (Chair).

\bibitem{Vtx-Review} ILC Vertex Chamber R\&D Review, to be held at the
ILC Workshop at Fermilab, October 2007.

  
  \bibitem{DOD} 4th Concept Detector Outline Document is available at the WWS-OC  website
    http://physics.uoregon.edu/~lc/wwstudy/concepts/.
     
\bibitem{LoI} ``Letter of Intent from the Fourth Detector (Ò4thÓ) Collaboration at the
International Linear Collider,'' available at {\tt www.4thconcept.org/4LoI.pdf}, and appendices on this same website.
     
   \bibitem{pisa}  ''Improving Spatial Resolution and Particle Identification'', G.F. 
   Tassielli, F. Grancagnolo, S. Spagnolo, 10th Pisa Meeting on Advanced Detectors,
   21-27 May 2006.

\bibitem{SPSC}  {\tt http://highenergy.phys.ttu.edu/dream/resources/proposals/SPSC2013.pdf}

\bibitem{dream-ha}  ''Hadron and Jet Detection with a Dual-Readout Calorimeter'',
    N. Akchurin, {\it et al.},
    {\em Nucl. Instr. Meths.} {\bf A 537} (2005) 537-561.

\bibitem{dream-n}   ``Neutron Signals for Dual-Readout Calori\-metry'', 
         Akchurin, N., {\it et al.},    {\it NIM} {\bf A598}(2009) 422-431; and, ``Measurement of the Contribution of Neutrons to Hadron Calorimeter Signals,''  
              						Akchurin, N., {\it et al.}, 	 {\it NIM} {\bf A581} (2007) 643-650.
  
 
 \bibitem{RD52-EM} ``The electromagnetic performance of the RD52 fiber calorimeter,'' submitted to {\it NIM}.
 
 \bibitem{RD52-pID}  ``Particle identification in the longitudinally unsegmented RD52 fiber calorimeter,'' submitted to {\it NIM}.
 
 \bibitem{Mik}  ''A Few Comments on the Status of Detectors for ILC'', Alexander Mikhailichenko, 
           CLNS 06/1951, 15 Jan 2006. 
 
 \bibitem{franco-mu-val}  ''The Muon System of the 4th Concept Detector'',   F.Grancagnolo,  ILC Workshop - ECFA and GLD Joint Meeting, Valencia, Spain, Nov. 5-13, 2006.
 
 \bibitem{telnov} V. Telnov, $\gamma$-$\gamma$ physics.
  
\bibitem{dream-e} ''Electron Detection with a Dual-Readout Calorimeter'',
    N. Akchurin, {\it et al.},
   {\em Nucl. Instr. Meths.} {\bf A 536}  (2005) 29-51.
 
\bibitem{dream-mu}  ''Muon Detection with a Dual-Readout Calorimeter'',
    N. Akchurin, {\it et al.},
    {\em Nucl. Instr. Meths.} {\bf A 533} (2004) 305-321.  

 \bibitem{spacal} Acosta, D., {\it et al.}, {\it NIM}, {\bf A302} (1991) 36.
 
\end{thebibliography}
\end{document}